\documentclass[aps,prl,10pt]{revtex4-1}
\usepackage[latin9]{inputenc}
\usepackage{amsmath}
\usepackage{amssymb}
\usepackage{graphicx}
\usepackage{bbm}

\begin{document}

\title{Non-commutativity measure of quantum discord}

\author{Yu Guo\footnote{Correspondence to guoyu3@aliyun.com}}

\affiliation{
School of Mathematics and Computer Science, Shanxi Datong University, Datong, Shanxi 037009, China
}

\begin{abstract}
Quantum discord is a manifestation of quantum correlations
due to non-commutativity rather than entanglement.
Two measures of quantum discord by the amount of non-commutativity
via the trace norm and the Hilbert-Schmidt norm respectively are proposed in this paper.
These two measures can be calculated easily for any state with arbitrary dimension.
It is shown by several examples that these measures can reflect
the amount of the original quantum discord.
\end{abstract}

\maketitle



\bigskip

\noindent
{\large \bf Introduction}

\noindent The characterization of quantum correlations in composite quantum states
is of great importance in quantum information theory
\cite{Nielsen,Horodecki1,Guhne,Ollivier,Henderson,Luo1}.
It has been shown that there are quantum
correlations that may arise without entanglement, such as quantum
discord (QD) \cite{Ollivier}, measurement-induced nonlocality
(MIN)\cite{Luo1}, quantum deficit \cite{Oppenheim},
quantum correlation induced by unbiased bases~\cite{Wu2014,Guowu2014} and
quantum correlation derived from the distance between the reduced states~\cite{Guo2014ijtp}, etc.
Among them, quantum discord has aroused great interest
in the past decade
~\cite{Shanchuanjia,Zurek2003,Werlang,Girolami,yusixia2011,Dujiangfeng,Chuan2012,Modi2012,
Paula,Huang2013,Streltsov2013,Fanheng2013,Libopla,Spehner,
Jakobczyk,Yan,Yang,Brodutch,Zhangchengjie,Yusixia}.
It is more robust against the effects of decoherence~\cite{Werlang}
and can be a resource in quantum computation
\cite{Datta,Brodutch2013}, quantum key distribution \cite{Suxiaolong2014}
remote state preparation \cite{Dakic12,Giorgi2013pra} and quantum cryptography \cite{Pirandola}.

Quantum discord is initially introduced by Ollivier and Zurek
\cite{Ollivier} and by Henderson and Vedral \cite{Henderson}.
The idea is to measure the discrepancy
between two natural yet different quantum analogs
of the classical mutual information.
For a state $\rho$ of a bipartite system A+B described by Hilbert space $H_a\otimes
H_b$, the quantum discord of $\rho$ (up to part B) is defined by
\begin{eqnarray}
D(\rho):=\min_{\Pi^b}\{I(\rho)-I(\rho|\Pi^b)\},\label{qd}
\end{eqnarray}
where, the minimum is taken over all local von Neumann measurements
$\Pi^b$,
$I(\rho):=S(\rho_a)+S(\rho_b)-S(\rho)$
is interpreted as the quantum mutual information,
$S(\rho):=-{\rm Tr}(\rho\log\rho)$
is the von Neumann entropy,
$I(\rho|\Pi^b)\}:=S(\rho_a)-S(\rho|\Pi^b)$,
$S(\rho|\Pi^b):=\sum_kp_kS(\rho_k)$,
and
$\rho_k=\frac{1}{p_k}( I_a\otimes\Pi_k^b)\rho(I_a\otimes\Pi_k^b )$
with $p_k={\rm Tr}[(I_a\otimes\Pi_k^b)\rho(I_a\otimes\Pi_k^b)]$,
$k=1$, 2, $\dots$, $\dim H_b$.
Calculation of quantum
discord given by Eq.~(\ref{qd}) in general is NP-complete since
it requires an optimization procedure over the set of all
measurements on subsystem B \cite{Huangyichen}. Analytical expressions
are known only for certain classes of
states~\cite{yusixia2011,Dujiangfeng,Huang2013,Libo,Dillenschneider,
Sarandy,Adesso,Giorda,Ali,chitambar,Luo08}.
Consequently, different versions (or measures) of quantum
discord have been proposed~\cite{Paula,Spehner,Jakobczyk,Brodutch2010,Dakic}:
the discord-like quantities in~\cite{Brodutch2010},
the geometric measure~\cite{Dakic}, the Bures distance measure~\cite{Spehner}
and the trace norm geometric measure ~\cite{Paula}, etc.
Unfortunately, all of theses measures are difficult to compute since they also need the
minimization or maximization scenario.

Let $\{|i_a\rangle\}$ be an orthonormal basis of $H_a$. Then any state
$\rho$ acting on $H_a\otimes H_b$ can be represented by
\begin{eqnarray}
\rho=\sum_{i,j}E_{ij}\otimes B_{ij},\label{staterepresentation}
\end{eqnarray}
where $E_{ij}=|i_a\rangle\langle j_a|$ and
$B_{ij}={\rm Tr}_a(|j_a\rangle\langle i_a|\otimes \mathbbm{1}_b\rho)$.
That is, assume that Alice and Bob share a state $\rho$, if Alice take an `operation'
\begin{eqnarray}
\Theta_{ij}:\rho\mapsto|j_a\rangle\langle i_a|\otimes \mathbbm{1}_b\rho
\end{eqnarray}
on her part,
then Bob obtains the local operator $B_{ij}$
(Note here that, the `operation' $\Theta_{ij}$ is not the
usual quantum operation
which admits the Kraus sum respresentation).
Quantum discord is from non-commutativity: $D(\rho)=0$ if and only if
$B_{ij}$s are mutually commuting normal operators~\cite{GuoHou,Dakic}.
It follows that the non-commutativity of the local operators $B_{ij}$s implies
$\rho$ contains quantum discord.
The central aim of this article is to show that, for any given state
written as in Eq.~(\ref{staterepresentation}), its quantum discord can be measured
by the amount of non-commutativity of the local operators, $B_{ij}$s.
In the following, we propose our approach: the non-commutativity measures.
We present two measures: the trace norm measure and the Hilbert-Schmidt norm one.
Both of them can be calculated for any state directly via the
Lie product of the local operators.
We then analyze our quantities for the Werner state, the isotropic state
and the Bell-diagonal state in which the original quantum discord have been calculated.
By comparing our quantities with the original one, we find that
our quantities can quantify quantum discord roughly for these states.

\bigskip

\noindent
{\large \bf Results}

\noindent
{\bf The amount of non-commutativity.} Let
$X$ and $Y$ be arbitrarily given operators on
some Hilbert space. Then $[X,Y]=XY-YX=0$
if and only if $\|[X,Y]\|=0$, $\|\cdot\|$ is any norm defined on the operator
space.
That is, $\|[X,Y]\|\neq0$ implies the non-commutativity of $X$ and $Y$.
In general, $\|[X,Y]\|$ reflects the amount of the non-commutativity of $X$ and $Y$.
Furthermore, for a set of operators $\Gamma=\{A_i: 1\leq i\leq n\}$,
the total non-commutativity of $\Gamma$ can be defined by
\begin{eqnarray}
N(\Gamma):=\sum\limits_{i<j}\|[A_{i},A_{j}]\|.\label{definitionofnoncommutativity}
\end{eqnarray}
In Ref.~\cite{Fei2014srep}, $N(\Gamma)$ is used for measure the
`quantumness' of a quantum ensemble $\Gamma$ when $\|\cdot\|$ is
the trace norm $\|\cdot\|_{\rm Tr}$, i.e., $\|A\|_{\rm Tr}={\rm Tr}\sqrt{A^\dag A}$.
We remark here that any norm can be used for quantifying the amount.
It is a natural way that, for any state as in Eq.~(\ref{staterepresentation}),
the amount of its non-commutativity
can be considered as the total non-commutativity of $\{B_{ij}\}$, $N(\{B_{ij}\})$.

\medskip

\noindent
{\bf Non-commutativity measure of quantum discord. } Let
$\rho=\sum_{i,j}E_{ij}\otimes B_{ij}$ be a state acting on $H_a\otimes H_b$
as in Eq.~(\ref{staterepresentation}).
We define a measure of QD for $\rho$ by
\begin{eqnarray}
D_N(\rho):=\sum\limits_{i\leq k,j\leq l}\|[B_{ij},B_{kl}]\|_{\rm Tr}
+\sum\limits_{i<k,l<j}\|[B_{ij},B_{kl}]\|_{\rm Tr}.\label{definitionofN}
\end{eqnarray}
Similarly, we can define
\begin{eqnarray}
D'_N(\rho):=\sum\limits_{i\leq k,j\leq l}\|[B_{ij},B_{kl}]\|_2
+\sum\limits_{i<k,l<j}\|[B_{ij},B_{kl}]\|_2,\label{definitionofN'}
\end{eqnarray}
where $\|\cdot\|_2$ denotes the Hilbert-Schmidt norm, i.e., $\|A\|_2=\sqrt{{\rm Tr}(A^\dag A)}$.
That is, if Alice takes $\Theta_{ij}$s on her part, $1\leq i,j\leq \dim H_a$,
then Bob can calculate the amount of non-commutativity through the reduced operators $B_{ij}$s.
By definition, it is obvious that i) $D_N(\rho)\geq0$, $D'_N(\rho)\geq0$, both $D_N$ and $D'_N$
vanish only for the zero quantum discord states, i.e., $D_N(\rho)=D'_N(\rho)=0$ iff $D(\rho)=0$;
ii) both $D_N$ and $D'_N$ are invariant under the local unitary
operations as that of the quantum discord, i.e., $D_N(\rho)=D_N(U_a\otimes U_b\rho U_a^\dag\otimes U_b^\dag)$
and $D'_N(\rho)=D'_N(U_a\otimes U_b\rho U_a^\dag\otimes U_b^\dag)$
for any unitary operator $U_{a/b}$ acting on $H_{a/b}$ (this implies that $D_N$
and $D'_{N}$ are independent on the choice of the local orthonormal bases
: if $\rho=\sum_{i,j}E_{ij}\otimes B_{ij}$ with respect to the local orthonormal
basis $\{|i_a\rangle|j_b\rangle\}$ and $\rho=\sum_{i,j}E_{ij}'\otimes B_{ij}'$
 with respect to another local orthonormal basis $\{|i_a'\rangle|j_b'\rangle\}$,
 then $E_{ij}'=U_aE_{ij}U_a^\dag$ and $B_{ij}'=U_bB_{ij}U_b^\dag$
 for some local unitary operators $U_a$ and $U_b$);
iii) $D_N(\rho)\geq D'_N(\rho)$ for any $\rho$.
By the definitions, it is clear that both $D_N$ and $D'_N$ can be easily calculated for any state.

Let $|\psi\rangle$ be a pure state with
Schmidt decomposition $|\psi\rangle=\sum_k\lambda_k|k_a\rangle|k_b\rangle$.
Then
\begin{eqnarray}
D_N(|\psi\rangle\langle\psi|)
&=&2\sum\limits_{i,j}\lambda_i\lambda_j(\sum\limits_{(k,l)\in\Omega}\lambda_k\lambda_l),\label{pure1}\\
D'_N(|\psi\rangle\langle\psi|)
&=&2\sum\limits_{i,j}\lambda_i\lambda_j(\sum\limits_{(k,l)\in\Omega'}\lambda_k\lambda_l)
+\sqrt{2},\label{pure2}
\end{eqnarray}
where $\Omega=\{(k,l):{\rm either}\ i< k\leq j\leq l\ {\rm or}\ k=i\ {\rm and}\ l=j\ {\rm if}\ i<j; i\leq k<l\ {\rm if}\ i=j\}$,
$\Omega'=\{(k,l):i< k\leq j\leq l\ {\rm if}\ i<j; i\leq k<l\ {\rm if}\ i=j\}$.
Therefore, $D_N(|\psi\rangle\langle\psi|)=0$ (or $D'_N(|\psi\rangle\langle\psi|)=0$)
if and only if $|\psi\rangle$ is separable.
For the maximally entangled state
$|\Psi^+\rangle=\frac{1}{\sqrt{d}}\sum_i|i_a\rangle|i_b\rangle$ in a
$d\otimes d$ system, it is straightforward that
$D_N(|\Psi^+\rangle\langle\Psi^+|)=\frac{3}{2}$ whenever $d=2$,
$\frac{8}{3}$ whenever $d=3$ and $4$ whenever $d=4$,
$D'_N(|\Psi^+\rangle\langle\Psi^+|)=1+\frac{\sqrt{2}}{4}$ whenever $d=2$, $2+\sqrt{2}$
whenever $d=3$ and $\frac{13}{4}+\frac{3\sqrt{2}}{8}$ whenever $d=4$.
$D_N$ and $D'_N$ reach the maximum values only on the maximally entangled one.

It is worth mentioning here that both $D_N$ and $D'_N$ are defined without measurement,
so the way we used is far different from the original quantum discord
and other quantum correlations (note that all the measures of the quantum correlations
proposed now are defined by some distance between the state and the post state after some
measurement). In addition, it is clear that $D_N(\rho)$ and $D'_N(\rho)$ are continuous
functions of $\rho$ since both the trace norm and Hilbert-Schmidt norm are continuous.
In \cite{Brodutch}, a set of criteria for measures of correlations are
introduced: (1) necessary conditions ((1-a)-(1-e)), (2) reasonable properties ((2-a)-(2-c)),
and (3) debatable criteria ((3-a)-(3-d)).
One can easily check that our quantity meets all the necessary conditions as a
measure of quantum correlation proposed in \cite{Brodutch}
(note that the condition (1-d) in \cite{Brodutch} is invalid for $D_N(\rho)$ and $D'_N(\rho)$).
The continuity of $D_N$ and $D'_N$ meets the reasonable property (2-a) (note: (2-b)
and (2-c) are invalid since these two conditions are associated with measurement-induced correlation).
(\ref{pure1}) and (\ref{pure2}) guarantee the debatable property (3-a).
(3-c) and (3-d) are not satisfied as that of the original quantum discord while
(3-b) is invalid for $D_N$ and $D'_N$.
That is, all the associated conditions that satisfied by the original quantum discord are met by our quantities.
From this perspective, $D_N$ and $D'_N$ are well-defined measures as that of the original quantum discord.

\medskip

\noindent
{\bf Comparing with the original quantum discord. }
In what follows, we compare the non-commutativity measures $D_N$ and $D'_N$ with quantum discord $D$
for several classes of well-known states and plot the level surfaces for the
Bell-diagonal states.
These examples will show that $D_N$ and $D'_N$ reflect the amount of quantum discord
roughly: $D_N$ and $D'_N$ increase (resp. decrease) if and only if $D$ increase (resp. decrease)
for almost all these states (see Figs. 1-3).  $D_N\geq D$ and $D'_N\geq D$ for almost all these states while
there do exist states such that $D_N< D$ and $D'_N< D$ (see Fig.~\ref{twoqubitstate} (a-b)).
In addition, $D_N$ and $D'_N$ characterize quantum discord in a more large scale than that of $D$ roughly.
For the two-qubit pure state $|\psi\rangle=\sum_k\lambda_k|k_a\rangle|k_b\rangle$, we can also calculate that
$D_N(|\psi\rangle\langle\psi|)>D(|\psi\rangle\langle\psi|)$ whenever $\lambda_1>a$ with $a\approx0.3841$
while $D_N(|\psi\rangle\langle\psi|)<D(|\psi\rangle\langle\psi|)$ whenever $\lambda_1<a$
and $D'_N(|\psi\rangle\langle\psi|)>D(|\psi\rangle\langle\psi|)$ whenever $\lambda_1>b$ with $b\approx0.4279$
while $D'_N(|\psi\rangle\langle\psi|)<D(|\psi\rangle\langle\psi|)$ whenever $\lambda_1<b$.

\begin{figure}
\centering\includegraphics[width=10cm]{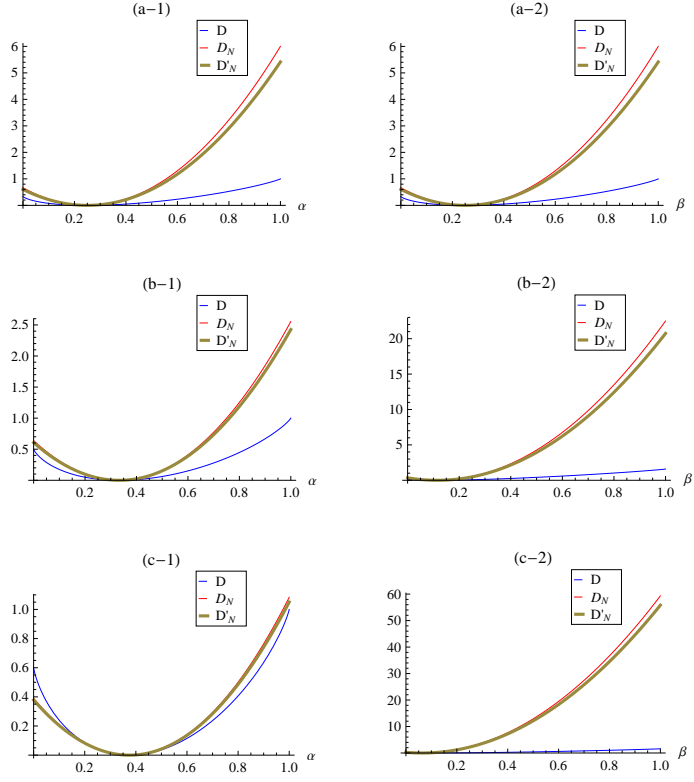}
\caption{\label{wernerstate}(color online). The measures $D$,
$D_N$ and $D'_N$ as functions of $\alpha$ for the Werner state when (a-1) $d=2$, (b-1) $d=3$  and (c-1) $d=4$,
and that of the isotropic state when (a-2) $d=2$, (b-2) $d=3$  and (c-2) $d=4$.
For both the Werner state and the isotropic state, $D_N$ and $D'_N$ are monotonic functions of $D$.}
\end{figure}

\noindent
{\bf Werner states.} The Werner states of a $d\otimes d$ dimensional
system admit the form\cite{Wer89},
\begin{eqnarray}
\rho_w=\frac{2(1-\alpha)}{d(d+1)}\Pi^++\frac{2\alpha}{d(d-1)}\Pi^-,
\quad \alpha\in[0,1], \label{p}
\end{eqnarray}
where $\Pi^+=\frac{1}{2}(I+F)$ and $\Pi^-=\frac{1}{2}(I-F)$ are projectors onto the
symmetric and antisymmetric subspace of $\mathbb{C}^d\otimes \mathbb{C}^d$ respectively,
$F=\sum_{i,j}|i_a\rangle\langle j_a|\otimes|j_b\rangle\langle i_b|$
is the swap operator.
Then
\begin{eqnarray}
D_N(\rho_w)=
\begin{cases}
\frac{2}{3}(1-4\alpha)^2,&  d=2,\\
\frac{23}{36}(1-3\alpha)^2,& d=3,\\
\frac{13}{300}(3-8\alpha)^2,&  d=4,
\end{cases}
\end{eqnarray}
and
\begin{eqnarray}
D'_N(\rho_w)=
\begin{cases}
\frac{4+\sqrt{2}}{9}(1-4\alpha)^2,&  d=2,\\
\frac{19+2\sqrt{2}}{36}(1-3\alpha)^2,& d=3,\\
\frac{35+2\sqrt{2}}{900}(3-8\alpha)^2,&  d=4.
\end{cases}
\end{eqnarray}
The three measures of quantum correlation,
i.e., $D_N$, $D'_N$ and $D$, are illustrated in
(a-1), (b-1) and (c-1) in Fig.~\ref{wernerstate}
for comparison, which reveals that the curves for
$D_N$ and $D'_N$ have the same tendencies as that of $D$.

\noindent
{\bf Isotropic states.}  For the $d\otimes d$ isotropic state
\begin{equation}
\rho_{is}=\frac{1}{d^2-1}((1-\beta)I+(d^2\beta-1)P^+),\quad \beta\in[0,1],
\label{Isotropicstates1}
\end{equation}
where $P^+=\frac{1}{d}\sum_{i,j}|i_a\rangle\langle j_a|\otimes|i_b\rangle\langle j_b|$
is the maximally entangled pure state in $\mathbb{C}^d\otimes \mathbb{C}^d$.
Then
\begin{equation}
D_N(\rho_{is})=
\begin{cases}
\frac{2}{3}(1-4\beta)^2,&  d=2,\\
\frac{3}{16}|1-9\beta|(|1-9\beta|+|1-8\beta)|),&  d=3,\\
|1-16\beta|(\frac{4}{25}|1-16\beta|+\frac{1}{9}|(1-15\beta|),& d=4
\end{cases}
\end{equation}
and
\begin{equation}
D'_N(\rho_{is})=
\begin{cases}
\frac{4+\sqrt{2}}{9}(1-4\beta)^2,&  d=2,\\
|1-9\beta|(\frac{6+3\sqrt{2}}{64}|1-9\beta|+\frac{3}{16}|1-8\beta)|),&  d=3,\\
|1-16\beta|(\frac{8+2\sqrt{2}}{75}|1-16\beta|+\frac{1}{9}|(1-15\beta|),& d=4
\end{cases}
\end{equation}
The three measures of quantum correlation,
i.e., $D_N$, $D'_N$ and $D$, are illustrated in
(a-2), (b-2) and (c-2) in Fig.~\ref{wernerstate}
for comparison.
We see from this figure that the curves for $D_N$ and $D'_N$ have the same tendencies as that of $D$.
It also implies that i) for both the Werner states and the isotropic states, $D_N$ and $D'_N$
are close to each other, ii) $D$ is close to $D_N$ and $D'_N$
with increasing of the dimension $d$ for the Werner states,
which in contrast to that of the isotropic states.
\begin{figure}
\centering \includegraphics[width=10cm]{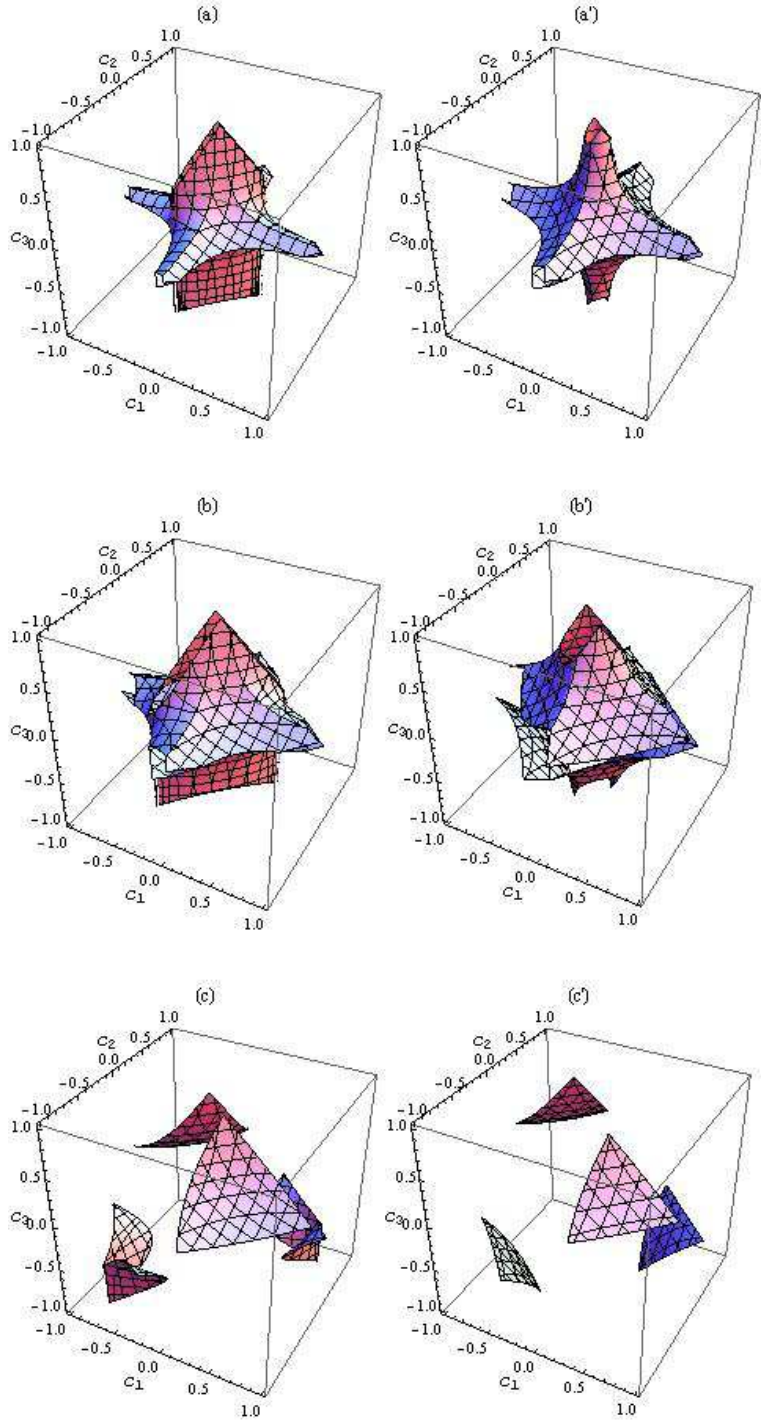}
\caption{\label{levelsurfaces1}(color online).
The surfaces of constant $D_N$ and $D'_N$ as functions of $c_1$, $c_2$ and $c_3$ for: (a) $D_N=0.05$,
(b) $D_N=0.1$ and (c) $D_N=0.3$;
(a$'$) $D'_N=0.05$,
(b$'$) $D'_N=0.1$ and (c$'$) $D'_N=0.3$.}
\end{figure}

\noindent
{\bf Bell-diagonal states.}  The Bell-diagonal states for two-qubits can be written as
\begin{equation}
\sigma_{ab}=\frac{1}{4} (I_2\otimes I_2+\sum_{j=1}^{3}c_{j}\sigma_{j}\otimes\sigma_{j} )
=\sum_{a,b}\lambda_{ab}|\beta_{ab}\rangle\langle\beta_{ab}|,
\label{twoqubitsymmtricstatessimplified}
\end{equation}
where the $\sigma_j$s are Pauli operators,
$\{|\beta_{ab}\rangle\}$ are four Bell states
$|\beta_{ab}\rangle\equiv\frac{1}{\sqrt{2}}(|0,b\rangle+(-1)^a|1,1\oplus b\rangle)$.
Then
\begin{eqnarray}
D_N\sigma_{ab})&=&\frac{1}{2}|c_1c_2|+\frac{|c_3|}{2}(|c_1-c_2|+|c_1+c_2|),\\
D'_N(\sigma_{ab})&=&\frac{1}{2\sqrt{2}}|c_1c_2|+\frac{|c_3|}{\sqrt{2}}\sqrt{c_1^2+c_2^2}.
\end{eqnarray}
In Fig.~\ref{levelsurfaces1}, the level surfaces of $D_N$
and $D'_N$ are plotted respectively.
By comparing them with that of $D$
in Ref.~\cite{Lang}, we find that the trends of $D_N$ and $D'_N$ are roughly the same as that of $D$:
$D_N$ and $D'_N$ increase when $D$ increases roughly and vice versa.
(The geometry of the set of the Bell-diagonal states is a tetrahedron
with the four Bell states sit at the four vertices, the extreme points of tetrahedron
(i.e., $(-1,1,1)$, $(1,-1,1)$, $(1,1,-1)$
and $(-1,-1,-1)$), see Fig.~1 in Ref.~\cite{Lang} for detail.)

Especially, we consider
\begin{eqnarray}
\rho_1& =& \frac{1}{2} |\beta_{01}\rangle\langle\beta_{01}|
+\frac{p}{2} |\beta_{00}\rangle\langle\beta_{00}|
+\frac{1-p}{2} |\beta_{10}\rangle\langle\beta_{10}|,\\
\rho_2 &=& p |\beta_{11}\rangle\langle\beta_{11}|
+\frac{1-p}{2}(|\beta_{01}\rangle\langle\beta_{01}|
+ |\beta_{00}\rangle\langle\beta_{00}| ),\\
\rho_3 &=& p|\beta_{11}\rangle\langle\beta_{11}|
+(1-p)|\beta_{01}\rangle\langle\beta_{01}|~~~~~~~~~~
\end{eqnarray}
and
\begin{eqnarray}
\rho_4 = p|\beta_{10}\rangle\langle\beta_{10}|
+(1-p)|\beta_{01}\rangle\langle\beta_{01}|.
\end{eqnarray}
The three measures of quantum correlation,
i.e., $D_N$, $D'_N$ and $D$, are compared in Fig.~\ref{twoqubitstate}.
For $\rho_1$, $\rho_3$ and $\rho_4$, the variation trends of  $D_N$ and $D'_N$ coincide with that of $D$
while for $\rho_2$ the curves of $D_N$ and $D'_N$ have the same tendency as that of $D$ roughly.
In addition, one can see that i) $D_N$ and $D'_N$ can both lager than and smaller than $D$,
namely, there is no order relation between $D$ and the two previous measures,
ii)  while the behavior of both measures $D_N$ and $D'_N$ is quite similar,
they are quite different from that of $D$.

Going further, we can quantify the symmetric quantum discord, i.e.,
the quantum discord up to both part A and part B.
Let $\{|k_b\rangle\}$ be an orthonormal basis of $H_b$, then any $\rho$ acting on
$H_a\otimes H_b$ admits the form
\begin{eqnarray}
\rho=\sum_{i,j}E_{ij}\otimes B_{ij}=\sum_{k,l}A_{kl}\otimes F_{kl}
\end{eqnarray}
with $F_{kl}=|k_b\rangle\langle l_b|$.
Here, $A_{kl}={\rm Tr}_b(\mathbbm{1}_a\otimes|l_b\rangle\langle k_b|\rho)$ are local operators
on $H_a$.
Let
\begin{eqnarray}
\tilde{D}_N(\rho):&=&\sum\limits_{i\leq k,j\leq l}\|[B_{ij},B_{kl}]\|
+\sum\limits_{i<k,l<j}\|[B_{ij},B_{kl}]\|\nonumber\\
&&+\sum\limits_{p\leq s,q\leq t}\|[A_{pq},A_{st}]\|+\sum\limits_{p<s,t<q}\|[A_{pq},A_{st}]\|,
\end{eqnarray}
where $\|\cdot\|$ is the trace norm, or the Hilbert-Schmidt norm, or other norms.
Then i) $\tilde{D}_N(\rho)\geq 0$ and $\tilde{D}_N(\rho)=0$
if and only if it is a classical-classical state
($\rho$ is called a classical-classical state
if $\rho=\sum_{i,j}p_{ij}|i_a\rangle\langle i_a|\otimes |j_b\rangle\langle j_b|$ with
$p_{ij}\geq 0$ and $\sum_{i,j}p_{ij}=1$);
ii)  $\tilde{D}_N$ is invariant under the local unitary operations.
We can conclude that $\tilde{D}_N(\rho)$ quantifies the amount of the symmetric
quantum discord of $\rho$.
\begin{figure}
\centering \includegraphics[width=10cm]{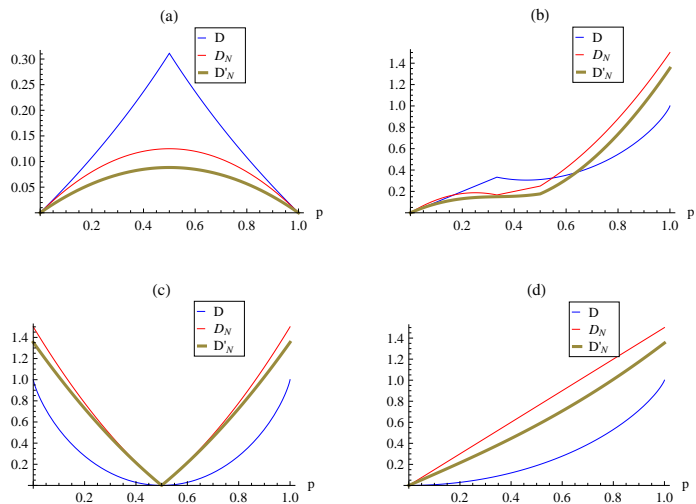}
\caption{\label{twoqubitstate}(color online).
The measures $D$, $D_N$ and $D'_N$ as functions of $p$ for (a) $\rho_1$, (b) $\rho_2$, (c) $\rho_3$ and
(d) $\rho_4$.}
\end{figure}

\bigskip

\noindent{\large \bf Discussion}

\noindent New measures of quantum discord has been proposed by
means of the amount of the non-commutativity
quantified by the trace norm and the Hilbert-Schmidt norm.
Our method provides two calculable measures of quantum discord from a new perspective:
unlike the original quantum discord and other
quantum correlations were induced by some measurement,
the two non-commutativity quantities
we presented were not defined via measurements.
Both of them can be calculated directly
for any state, avoiding the previous optimization procedure in calculation.
The nullities of our measures coincide with that of the original quantum discord
and they are invariant under local unitary operation as well.
The examples we analyzed indicate that,
when comparing our quantities with the original quantum discord,
although they are
different and even have large difference for some special states, the non-commutativity measures
reflect the original quantity
roughly overall.
We can conclude, to a certain extent, that our approach
can reflect the original quantum discord
for the set of states with arbitrary dimension.
On the other hand, the non-commutativity measures reflect quantum discord
in a larger scale than that of the original quantum discord,
we thus can use these measures to find quantum states with limited
quantum discord or the maximal discordant states
(especially for the states represented by one or two parameters), etc.

As usual, only the trace norm and the Hilbert-Schmidt norm are considered.
In fact we can also use the general operator norm or other norms in the definitions of $D_N$ and $D'_N$.
In addition, Fig.~\ref{levelsurfaces1} shows that the level surfaces of $D'_N$ are nearly symmetric
up to the four Bell states directions, which is very close to that of the quantum discord $D$
(the level surfaces of $D$ are symmetric
up to the four Bell states directions~\cite{Lang}).
Also note that the Hilbert-Schmidt norm is more easily calculated than the trace norm one,
we thus use the Hilbert-Schmidt norm measure
in general.

\medskip
\noindent
{\large \bf Acknowledgements}

\noindent

This work is supported by the National Natural Science Foundation
of China under Grant No. 11301312.

\medskip
\noindent
{\large \bf Additional Information}

\noindent{\bf Competing financial interests:} The authors declare no competing financial interests.


\begin{thebibliography}{10}

\bibitem{Nielsen} Nielsen, M.~A. \& Chuang, I.~L. \textit{Quantum Computatation and
Quantum Information}, (Cambridge University Press, Cambridge, England, 2000).

\bibitem{Horodecki1} Horodecki, R.,
Horodecki, P., Horodecki, M. \& Horodecki, K.
Quantum entanglement.
\emph{Rev. Mod. Phys.} \textbf{81}, 865 (2009).

\bibitem{Guhne} G\"{u}hne, O. \& T\'{o}th, G.
Entanglement detection.
\emph{Phys. Rep.} \textbf{474}, 1 (2009).

\bibitem{Ollivier} Ollivier, H. \& Zurek, W.~H.
Quantum discord: a measure of the quantumness of correlations.
\emph{Phys. Rev. Lett.} \textbf{88}, 017901 (2001).

\bibitem{Henderson} Henderson, L. \& Vedral, V.
Classical, quantum and total correlations.
\emph{J. Phys. A} \textbf{34}, 6899 (2001).

\bibitem{Luo1} Luo, S. \& Fu, S.
Measurement-induced nonlocality.
\emph{Phys. Rev. Lett.} \textbf{106}, 120401 (2011).

\bibitem{Oppenheim} Oppenheim, J., Horodecki, M., Horodecki, P. \& Horodecki, R.
Thermodynamical approach to quantifying quantum correlations.
\emph{Phys. Rev. Lett.} \textbf{89}, 180402 (2002).

\bibitem{Wu2014} Wu, S., Ma, Z., Chen, Z. \& Yu, S.
Reveal quantum correlation in complementary bases.
\emph{Sci. Rep.} \textbf{4}, 4036 (2014).

\bibitem{Guowu2014} Guo, Y. \& Wu, S.
Quantum correlation exists in any non-product state.
\emph{Sci. Rep.} \textbf{4}, 7179 (2014).

\bibitem{Guo2014ijtp} Guo, Y., Li, X., Li, B. \& Fan, H.
Quantum correlation induced by the average distance between the reduced states.
\emph{Int. J. Theor. Phys.} \textbf{54}(6), 2022-2030 (2015).

\bibitem{Shanchuanjia} Shan, C., Cheng, W., Liu, J., Cheng, Y. \& Liu, T.
Scaling of geometric quantum discord close to a topological phase transition.
\emph{Sci. Rep.} \textbf{4}, 4473 (2014).

\bibitem{Zurek2003} Zurek, W. H.
Quantum discord and Maxwell¡¯s demons.
\emph{Phys. Rev. A} \textbf{67}, 012320 (2003).

\bibitem{Werlang} Werlang, T., Souza, S., Fanchini, F. F. \& Villas Boas, C. J.
Robustness of quantum discord to sudden death.
\emph{Phys. Rev. A} \textbf{80}, 024103 (2009).

\bibitem{Girolami} Girolami, D. \& Adesso, G.
Quantum discord for general two-qubit states: Analytical progress.
\emph{Phys. Rev. A} \textbf{83}, 052108 (2011).

\bibitem{yusixia2011} Chen, Q., Zhang, C., Yu, S., Yi, X. X. \& Oh, C. H.
Quantum discord of two-qubit X states.
\emph{Phys. Rev. A} \textbf{84}, 042313 (2011).

\bibitem{Dujiangfeng} Shi, M., Yang, W., Jiang, F. \& Du, J.
Quantum discord of two-qubit rank-2 states.
\emph{J. Phys. A: Math. Theor.} \textbf{44}, 415304 (2011).

\bibitem{Chuan2012} Chuan, T. K. \emph{et al}.
Quantum discord bounds the amount of distributed entanglement.
\emph{Phys. Rev. Lett.} \textbf{109}, 070501 (2012).

\bibitem{Modi2012} Modi, K., Brodutch, A., Cable, H., Paterek, T. \& Vedral, V.
The classical-quantum boundary for correlations: Discord and related measures.
\emph{Rev. Mod. Phys.} \textbf{84}, 1655 (2012).

\bibitem{Paula} Paula, F. M., de Oliveira, T. R. \& Sarandy, M. S.
Geometric quantum discord through the Schatten 1-norm.
\emph{Phys. Rev. A} \textbf{87}, 064101 (2013).

\bibitem{Huang2013} Huang, Y.
Quantum discord for two-qubit X states: Analytical formula with very small worst-case error.
\emph{Phys. Rev. A} \textbf{88}, 014302 (2013).

\bibitem{Streltsov2013} Streltsov, A. \& Zurek, W. H.
Quantum discord cannot be shared.
\emph{Phys. Rev. Lett.} \textbf{111}, 040401 (2013).

\bibitem{Fanheng2013} Hu, M. L. \& Fan, H.
Upper bound and shareability of quantum discord based on entropic uncertainty relations.
\emph{Phys. Rev. A} \textbf{88}, 014105 (2013).

\bibitem{Libopla} Li, B., Chen, L. \& Fan, H.
Non-zero total correlation means non-zero quantum correlation.
\emph{Phys. Lett. A} \textbf{378}, 1249-1253 (2014).

\bibitem{Spehner} Spehner, D. \& Orszag, M.
Geometric quantum discord with Bures distance: the qubit case.
 \emph{J. Phys. A: Math. Theor.} \textbf{47}, 035302 (2014).

\bibitem{Jakobczyk} Jak\'{o}bczyk, L.
Spontaneous emission and quantum discord: Comparison of Hilbert-Schmidt and trace distance discord.
\emph{Phys. Lett. A} \textbf{378}, 3248-3253 (2014).

\bibitem{Yan} Yan, X. \& Zhang, B.
Collapse-revival of quantum discord and entanglement.
\emph{Ann. Phys.} \textbf{349}, 350-356 (2014).

\bibitem{Yang} Yang, X., Huang, G. \& Fang, M.
A study on quantum discord in Gaussian states.
\emph{Opt. Commun.} \textbf{341}, 91-96 (2015).

\bibitem{Brodutch} Brodutch A. \& Modi K. Criteria for measures of quantum correlations.
Quant. Inf. \& Comput. 12, 0721 (2012).

\bibitem{Zhangchengjie} Zhang, C. \emph{et al}.
Complete condition for nonzero quantum correlation in continuous varialbe systems.
New J. Phys. 17, 093007 (2015)

\bibitem{Yusixia} Yu, S., Zhang, C., Chen, Q. \& Oh, C. H.
Witnessing the quantum discord of all the unknown states.
arXiv:quant-ph/1102.4710.

\bibitem{Datta} Datta, A., Shaji, A. \& Caves, C.~M.
Quantum discord and the power of one qubit.
\emph{Phys. Rev. Lett.} \textbf{100}, 050502 (2008).

\bibitem{Brodutch2013} Brodutch, A.
Discord and quantum computational resources.
\emph{Phys. Rev. A} \textbf{88}, 022307 (2013).

\bibitem{Suxiaolong2014} Su, X.
Applying Gaussian quantum discord to quantum key distribution.
\emph{Chin. Sci. Bull.} \textbf{59}, 1083-1090 (2014).

\bibitem{Dakic12} Daki\'{c}, B. \emph{et al}.
Quantum discord as resource for remote state preparation.
\emph{Nature Phys.} \textbf{8}, 666 (2012).


\bibitem{Giorgi2013pra} Giorgi, G. L.
Quantum discord and remote state preparation.
\emph{Phys. Rev. A} \textbf{88}, 022315 (2013).

\bibitem{Pirandola} Pirandola, S.
Quantum discord as a resource for quantum cryptography.
\emph{Sci. Rep.} \textbf{4}, 6956 (2014).

\bibitem{Huangyichen} Huang, Y.
Computing quantum discord is NP-complete.
\emph{New J. Phys.} \textbf{16}, 033027 (2014).

\bibitem{Libo} Li, B., Wang, Z. X. \& Fei, S. M.
Quantum discord and geometry for a class of two-qubit states.
\emph{Phys. Rev. A} \textbf{83}, 022321 (2011).

\bibitem{Brodutch2010} Brodutch, A. \& Terno, D. R.
Quantum discord, local operations, and Maxwell's demons.
\emph{Phys. Rev. A} \textbf{81}, 062103 (2010).

\bibitem{Dillenschneider} Dillenschneider, R.
Quantum discord and quantum phase transition in spin chains.
\emph{Phys. Rev. B} \textbf{78}, 224413 (2008).

\bibitem{Sarandy} Sarandy, M. S.
Classical correlation and quantum discord in critical systems.
\emph{Phys. Rev. A} \textbf{80}, 022108 (2009).

\bibitem{Adesso} Adesso, G. \& Datta, A.
Quantum versus classical correlations in Gaussian states.
\emph{Phys. Rev. Lett.} \textbf{105}, 030501 (2010).

\bibitem{Giorda} Giorda, P. \&
Paris, M. G. A.
Gaussian quantum discord.
\emph{Phys. Rev. Lett.} \textbf{105}, 020503 (2010).

\bibitem{Ali} Ali, M., Rau, A. R. P. \& Alber, G.
Quantum discord for two-qubit X states.
\emph{Phys. Rev. A} \textbf{81}, 042105 (2010).

\bibitem{chitambar} Chitambar, E.
Quantum correlation in high-dimensional states of high symmetry.
\emph{Phys. Rev. A} \textbf{86}, 032110 (2012).

\bibitem{Luo08} Luo, S.
Quantum discord for two-qubit systems.
\emph{Phys. Rev. A} \textbf{77}, 042303 (2008).

\bibitem{Dakic} Daki\'{c}, B., Vedral, V. \& Brukner, \v{C}.
Necessary and sufficient condition for nonzero quantum discord.
\emph{Phys. Rev. Lett.} \textbf{105}, 190502 (2010).

\bibitem{GuoHou} Guo, Y. \& Hou, J.
A class of separable quantum states.
\emph{J. Phys. A: Math. Theor.} \textbf{45}, 505303 (2012).

\bibitem{Fei2014srep} Ma, T., Zhao, M., Wang, Y. \& Fei, S.
Non-commutativity and local indistinguishability of quantum state.
\emph{Sci. Rep.} \textbf{4}, 6336 (2014).

\bibitem{Wer89} Werner, R. F.
Quantum states with Einstein-Posolsky-Rosen correlations admitting a hidden-variable model.
\emph{Phys. Rev. A} \textbf{40}, 4277 (1989).


\bibitem{Lang} Lang, M. D. \& Caves, C. M.
Quantum discord and the geometry of Bell-diagonal states.
\emph{Phys. Rev. Lett.} \textbf{105}, 150501 (2010).


\end{thebibliography}
\end{document}